\documentclass[aps,prl, superscriptaddress, showpacs,preprint]{revtex4-1}

\usepackage[]{graphicx}% Include figure files
\usepackage{dcolumn}% Align table columns on decimal point
\usepackage{bm}% bold math
\usepackage{amsmath}
\usepackage{amssymb}
\usepackage{verbatim}
\usepackage{appendix}
\usepackage{float}

\usepackage{placeins}
\usepackage{color,soul}
\usepackage{braket}
\usepackage{amsmath}

\begin{document}

%\preprint{}

%\title{Antisymmetric Couplings Enable Spectroscopic Resolution of Energy Equivalent Enantiomers in Nuclear Magnetic Resonance Spectroscopy}
\title{Antisymmetric Couplings Enable Direct Observation of Chirality in Nuclear Magnetic Resonance Spectroscopy}


\author{Jonathan P. King}
\email[]{jpking@berkeley.edu}
\affiliation{Department of Chemistry, University of California, Berkeley, California 94720, USA}
\affiliation{Materials Sciences Division, Lawrence Berkeley National Laboratory, Berkeley, California 94720, USA}
\author{Tobias F. Sjolander}
\affiliation{Department of Chemistry, University of California, Berkeley, California 94720, USA}
\affiliation{Materials Sciences Division, Lawrence Berkeley National Laboratory, Berkeley, California 94720, USA}
\author{John W. Blanchard}
%\author{Alexander P. Pines}
\affiliation{Department of Chemistry, University of California, Berkeley, California 94720, USA}
\affiliation{Materials Sciences Division, Lawrence Berkeley National Laboratory, Berkeley, California 94720, USA}
\affiliation{Helmholtz-Institut Mainz, Germany}

%\thanks{}


\date{\today}

\begin{abstract}
Here we demonstrate that a term in the nuclear spin Hamiltonian, the antisymmetric \textit{J}-coupling, is fundamentally connected to molecular chirality. We propose and simulate a nuclear magnetic resonance (NMR) experiment to observe this interaction and differentiate between enantiomers without adding any additional chiral agent to the sample. The antisymmetric \textit{J}-coupling may be observed in the presence of molecular orientation by an external electric field. The opposite parity of the antisymmetric coupling tensor and the molecular electric dipole moment yields a sign change of the observed coupling between enantiomers. We show how this sign change influences the phase of the NMR spectrum and may be used to discriminate between enantiomers. %Furthermore, chiral signals may be obtained from racemic mixtures, which may be distinguished from achiral compounds. This possibility arises from the nonlinear nature of NMR, where the resultant signal of right- and left-handed molecules need not be equal in magnitue and opposite in sign. %These spectroscopic signals are analogous to optical rotation of light, but may be resolved into spectral peaks that carry chemical-specific information. To our knowledge, such an experiment would be the first to spectrally resolve chiral molecules in a racemic mixture.
\end{abstract}

%\pacs{76.60.-k, 76.30.Mi, 82.56.Na}
%\keywords{}

\maketitle

Molecular chirality is an extremely important yet often difficult to measure property. Chiral molecules form many of the basic building blocks of living organisms, such as L-amino acids and D-sugars. Because of this biochirality, the effect of chiral pharmaceuticals can differ greatly between enantiomers \cite{Hutt1996}, leading to great interest in enantioselective synthesis and methods to analyze chiral products \cite{Busch}. Molecular chirality is also of interest in the study of fundamental symmetries \cite{Barra1996}, where parity violation in molecules is predicted but has not yet been observed \cite{ANIE:ANIE200290005,C0CP01483D,doi:10.1146/annurev.physchem.58.032806.104511}.  The search for spectroscopic techniques to probe chirality is a field of great interest where many proposals and experiments may be found in the recent literature \cite{Yachmenev2016,patterson2013,pattersonreview,fischer}. Meanwhile, liquid-state NMR is the preeminent technique for chemical analysis owing to its generality and chemical specificity. The introduction of direct chiral detection to NMR, without the need to add derivatizing agents to the sample, would be a major step forward in the field of chiral analysis. Several proposals exist for direct observation of chirality via an electric-field-induced pseudoscalar term in the NMR Hamiltonian. These involve either the detection of an induced oscillating molecular electric dipole moment \cite{Buckingham2004,Buckingham2006,Buckingham2014,paramagnetic}  or a magnetic dipole signal induced via application of an electric field \cite{Walls2008,Walls2014,Harris2006}.

In this Letter, we present a method for detection of chirality using liquid-state zero-field NMR, taking advantage of the antisymmetric spin-spin coupling. By analyzing the symmetry properties of this interaction we first demonstrate the connection between the antisymmetric \textit{J}-coupling tensor and molecular chirality. We then propose and simulate a prototype experiment using established techniques and predict a signal amplitude comparable to previous zero-field NMR experiments. Although not yet observed, terms in the Hamiltonian arising from the antisymmetric \textit{J}-coupling are predicted in a wide array of molecules from symmetry considerations \cite{Buckingham1982,Buckingham1970}, perturbational calculations\cite{Robert1982,Andrew1968,Schneider1968} and from quantum chemical calculations \cite{Vaara2002,Bryce2000}. In addition to the connection to chirality made here, measurement of the antisymmetric \textit{J}-coupling has been suggested as a means to observe molecular parity violation \cite{Barra1996}, which is predicted to cause a first-order energy shift to the antisymmetric J-coupling.

The \textit{J}-coupling coupling Hamiltonian (in frequency units) for two spins may be written:

\begin{equation}
H=I_1\cdot J\cdot I_2.
\label{ham1}
\end{equation}

\noindent  $I_1$ and $I_2$ are vector spin operators, and $J$ contains the spatial coordinates. The total nuclear spin-spin coupling tensor may include terms from magnetic dipole-dipole coupling as well as the \textit{J}-coupling, but the antisymmetric terms we are interested in can only arise from the \textit{J}-coupling. This is because the \textit{J}-coupling involves indirect interactions through electron spins, which need not have local inversion symmetry while the magnetic dipole-dipole coupling is symmetric under inversions and is described by a rank-2 irreducible spherical tensor. The \textit{J} tensor describes the coupling between two angular momenta and may be decomposed into irreducible spherical tensor components up to rank-2:

\begin{equation}
J=J^{(0)}+J^{(1)}+J^{(2)},
\label{Jtensor}
\end{equation}

\noindent where the $J^{(1)}$ component is antisymmetric with respect to the spatial coordinates. Since $J^{(1)}$ is traceless, its components average to zero for unoriented molecules, and thus cannot be observed in isotropic liquids. Alignment techniques such as liquid crystals and stretched gels can revive rank-2 terms, but to observe rank-1 interactions molecular orientation is required (see Supplemental Material). Solid-state studies in principle could detect antisymmetric couplings, but in practice have insufficient resolution \cite{tin}. Here we propose molecular orientation by an applied electric field\cite{Buckingham1963,riley2000}.

$J^{(1)}$ has three independent quantities, $J^{(1)}_{\alpha\beta}$, where $\alpha,\beta=x, y, z$:

%\begin{equation}

%\label{matrix}
%\end{equation}
\begin{equation}
J^{(1)}=\left( \begin{array}{ccc}
0 & J_{xy} & J_{xz} \\
-J_{xy} & 0 & J_{yz} \\
-J_{xz} & -J_{yz} & 0 \end{array} \right).
\label{matrix}
\end{equation}

%\[ \left( \begin{array}{ccc}
%0 & J^{(1)}_{z} & J^{(1)}_{y} \\
%-J^{(1)}_{z} & 0 & J^{(1)}_{x} \\
%-J^{(1)}_{y} & -J^{(1)}_{x} & 0 \end{array} \right)\]

\noindent  It has been shown that the number of independent components of the $J$ tensor depend on the local symmetry of the spin pair, and that three independent components of $J^{(1)}$ exist only if the two nuclear sites have local C$_1$ symmetry \cite{Buckingham1982}, which includes all chiral molecules and meso compounds.

We now address the question: ``Given $J^{(1)}$ in a molecule-fixed coordinate system for a chiral molecule, how will it be different for the complementary enantiomer?" The transformation between enantiomers may be visualized as a reflection through a plane containing the molecular electric dipole vector (which we will call the z-direction). The electric dipole will remain unchanged, as required by the orienting field, while the xy-component of $J^{(1)}$ will acquire a negative sign ($J_{xy}\rightarrow J_{x(-y)}=-J_{xy}$) (Fig. 1). This sign change is a manifestation of the pseudovector nature of $J^{(1)}$. Since we are considering oriented liquids rapidly rotating around the z-axis, the time-averaged coupling tensor has only the single xy-component which is scaled by the degree of orientation (see Supplemental Material):

\begin{equation}
\overline{J^{(1)}}\propto\left( \begin{array}{ccc}
0 & \pm J_{xy} & 0 \\
\mp J_{xy} & 0 & 0 \\
0 & 0 & 0 \end{array} \right),
\end{equation}

\noindent where the constant of proportionality comes from the degree of orientation and the sign depends on the handedness of the molecule. A measurement of the sign of $J_{xy}$, which is equal to $J^{(1)}_{0}$ in the spherical basis, yields the chiral signal. We note that $\overline{J^{(1)}_0}$ is the projection of $J^{(1)}$ along the director axis, scaled by the degree of orientation. Since the orientation is defined by the molecular electric dipole moment, the measured quantity is proportional to $J^{(1)}\cdot E$, which gives a pseudoscalar term in the Hamiltonian. The remainder of this work outlines methods to observe the sign of $\overline{J^{(1)}_0}$.
%This matrix can be written more compactly in summation notation using the antisymmetric Levi-Civita tensor:

 \begin{figure}[h]
 \includegraphics[width=0.9\textwidth]{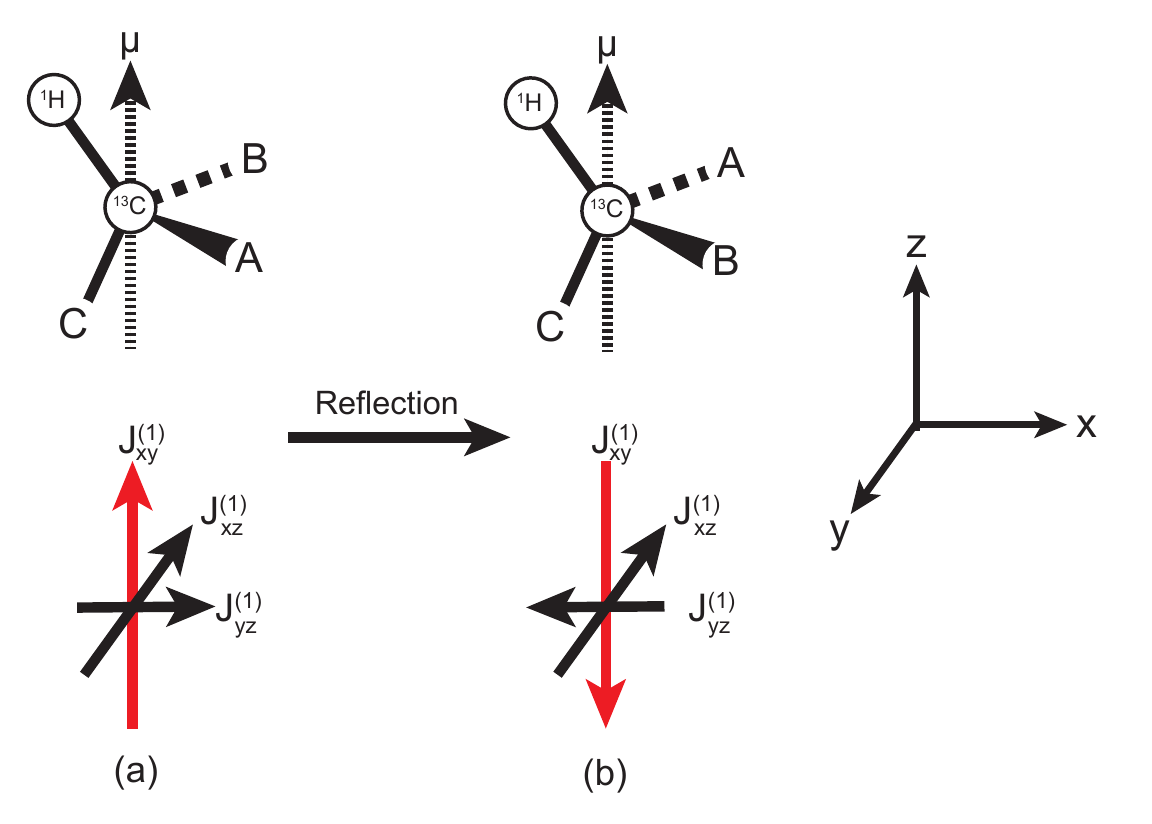}%
 \caption{\label{fig1}a) Schematic chiral molecule and $J^{(1)}$ tensor components for the $^1$H-$^{13}$C J-coupling. $\mu$ is the molecular electric dipole which is oriented along the z-axis by an applied electric field. b) After reflection through the xz plane, the electric dipole is unchanged, but the xy-component of $J^{(1)}$ has now gained a negative sign. In both situations the molecule is rapidly rotating around the z-axis, so only the xy (red) component of $J^{(1)}$ is observable.}
 \label{chiral}
 \end{figure}

 \begin{figure}[h]
 \includegraphics[width=0.9\textwidth]{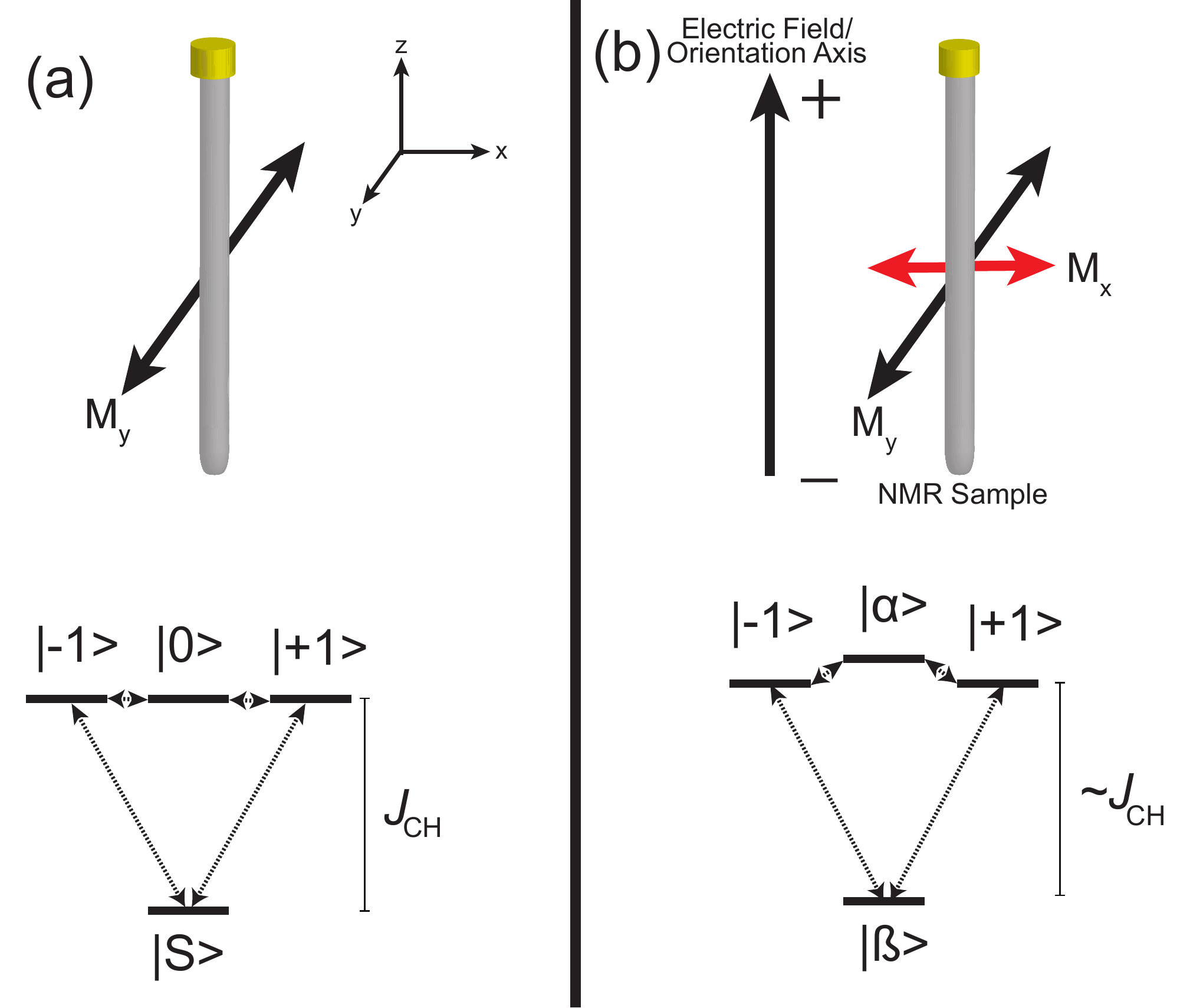}%
 \caption{\label{schematic}Schematic representation of a zero-field NMR experiment for chiral discrimination. a) Without electric field orientation, we have pure singlet and triplet states and no signal emerges. b) With electric field orientation, the singlet and $|0\rangle$ states mix and the coherences become observable as oscillating transverse magnetization along the x-axis. The choice of initial magnetization, electric field vector, and transverse detector vector define a handed coordinate system that yields chiral information in the sign of the detected signal. The $\alpha$ and $\beta$ states exhibit small energy shifts from $\overline{J^{(1)}_0}$ and residual dipolar couplings.}
 \label{schematic}
 \end{figure}

The antisymmetric \textit{J}-coupling Hamiltonian does not commute with the Zeeman Hamiltonian unless the spins have the same Larmor frequency. In typical high-field NMR an identical Larmor frequency would imply that the spins be chemically equivalent, which is incompatible with local C$_1$ symmetry. %However, if the Larmor frequency is zero, the full \textit{J}-coupling is observable. Recent work in zero-field NMR has demonstrated detection of small spin-spin couplings that would be suppressed in a high magnetic field \cite{Blanchard2015}.

We therefore propose an experiment based on zero-field NMR \cite{emagres} of two coupled unlike spin-$\frac{1}{2}$ nuclear spins in a chiral molecule oriented by an electric field. Zero-field NMR spectroscopy is the detection of NMR signals generated by spin-spin couplings in the absence of an applied magnetic field, thus giving an identical Larmor frequency of zero for all spins. The spin-spin coupling signals are generally in the sub-kHz regime and are detected through means such as atomic vapor cell magnetometers. In isotropic media, the signal corresponds to $J^{(0)}$. We show how an additional signal, whose sign and magnitude depend on $\overline{J^{(1)}_0}$, may be observed along an axis orthogonal to the usual signal. The general principle is that the three axes defined by the initial spin state, detector axis, and electric field form a coordinate system with a definite handedness that yields chiral information (Fig. \ref{schematic}).

We define bilinear spherical tensor operators for coupled spins $I_1$ and $I_2$:

\begin{equation}
T^{(q)}_k=\sum^1_{k_1=-1}\sum^1_{k_2=-1}C^{q,k}_{1,k_1,1,k_2}I^{(1)}_{k_1}I^{(1)}_{k_2},
\label{operators}
\end{equation} 

\noindent where $C^{q,k}_{q_1,k_1,q_2,k_2}$ are Clebsch-Gordan coefficients. With these definitions, we can write the time-averaged coupling Hamiltonian in a more useful form:

\begin{equation}
 H=I_1\cdot\overline{J}\cdot I_2=\sum_{q=0}^2\sum_{k=-q}^q (-1)^k \overline{J^{(q)}_k} T^{(q)}_{-k}.
\label{ham2}
\end{equation} 

\noindent However, since there is rapid motion around the z-axis, this reduces to (see Supplemental Material):

\begin{equation}
H=\sum_{q=0}^2\overline{J^{(q)}_0} T^{(q)}_0,
\label{ham3}
\end{equation}

\noindent where the sign of $\overline{J^{(1)}_0}$ depends on the handedness of the molecule as shown above. As will be shown in the simulations, the presence of rank-2 couplings (either from the rank-2 \textit{J} tensor or residual dipolar couplings) give small first-order frequency shifts \cite{Blanchard2015} while $\overline{J^{(1)}_0}$ gives a small second-order shift (Fig. 2b) but none of these interfere with chiral discrimination. We therefore neglect rank-2 terms in the analytical discussion and write the relevant Hamiltonian:

\begin{equation}
H=J^{(0)}_0T^{(0)}_0+\overline{J^{(1)}_0} T^{(1)}_0.
\label{ham4}
\end{equation}

In our NMR experiment, the observable quantity is the magnetization along the detector axis, chosen here to be the x-axis, and represented by the operator $M_x$. Note that for this experiment there will also be a large achiral signal along the y-axis. In the spherical basis, the observable operator is:

\begin{equation}
M_x=\sum_i\gamma_iI_{x,i}=\sum_i\gamma_i(I^{(1)}_{+1,i}+I^{(1)}_{-1,i})
\end{equation}

\noindent and the observable signal which is given by:

\begin{equation}
\langle M_x(t)\rangle(t)=\mathrm{Tr}\lbrace M^\dagger_x\rho(t)\rbrace.
\label{signal}
\end{equation}

\noindent  In the absence of $J^{(1)}_0$, the energy eigenstates are three triplet states $|+1\rangle,|0\rangle,|-1\rangle$ and a singlet $|S\rangle$ (Fig. 2a). We consider $\overline{J^{(1)}_0}$ as a perturbation, which is valid when $\overline{\vert J^{(1)}_0}\vert<<\vert J^{(0)}\vert$ as is the case for weakly oriented molecules. We note, however, that weak orientation is not a necessary condition for chiral discrimination. While the $|+1\rangle$ and $|-1\rangle$ states are not affected by the presence of $\overline{J^{(1)}_0}$, the $|0\rangle$ and $|S\rangle$ states are mixed so that the first-order eigenstates are:

\begin{equation}
|\alpha\rangle=(|S\rangle-i\frac{\overline{J^{(1)}_0}}{2J^{(0)}}|0\rangle)\frac{1}{N}
\label{}
\end{equation}

and

\begin{equation}
|\beta\rangle=(|0\rangle-i\frac{\overline{J^{(1)}_0}}{2J^{(0)}}|S\rangle)\frac{1}{N} 
\label{}
\end{equation}

\noindent where $N$ is a normalization factor. (See Fig. 2b with energy shifts from $\overline{J^{(1)}_0}$ and residual dipolar couplings) The presence of a term linear in $\overline{J^{(1)}_0}$ enables the creation of observable signals also linear in $\overline{J^{(1)}_0}$. We choose an experimentally realizable initial condition containing coherences involving $|\alpha\rangle$ and $|\beta\rangle$ (arrows in Fig. 2b). Our chosen initial density operator is

\begin{equation}
\rho(0)=\frac{1}{4}+\frac{B_\mathrm{p}\hbar}{4kT}\left[\frac{\gamma_1+\gamma_2}{2}(-I_{y,1}+I_{y,2})+\frac{\gamma_1-\gamma_2}{2}(-I_{y,1}-I_{y,2})\right],
\end{equation}

\noindent which corresponds to prepolarization of spins in a field $B_\mathrm{p}$ along the y-axis at temperature $T$ followed by inversion of spin 1. $k$ is the Boltzmann constant. This initial condition gives the following predicted signal in the zero-field NMR experiment (See Supplemental Material for a full calculation of matrix elements and coherence amplitudes):
 
  \begin{equation}
\langle M_x(t)\rangle=\frac{B_\mathrm{p}\hbar\gamma_1\gamma_2\overline{J^{(1)}_0}}{kTN^2J^{(0)}}\left[\cos(\omega_\alpha t)-\cos(\omega_\beta t)\right],
\label{signal}
\end{equation}

\noindent where $\omega_\alpha=\frac{E_\alpha-E_{\pm1}}{\hbar}$ and $\omega_\beta=\frac{E_\beta-E_{\pm1}}{\hbar}$.

In our simulations, we consider the case of a chiral molecule with two spin-$\frac{1}{2}$ nuclei ($^{13}$C and $^1$H) with $J^{(0)}=100$ Hz, typical of a one-bond $^1$H-$^{13}$C \textit{J}-coupling. Since antisymmetric \textit{J}-couplings are similar in magnitude to the isotropic rank-0 term \cite{Vaara2002}, we assume a residual $\overline{J^{(1)}_0}$ of 1 Hz, corresponding to a orientational order parameter of $\sim10^{-2}$, typical of electric-field orientation experiments\cite{riley2000}. For a one-bond $^{13}$C-$^1$H coupling with this degree of orientation, the alignment induced by the electric field results in a residual dipolar coupling of $\sim0.7$ Hz (Supplemental Material) which results in small frequenecy shifts.. Oscillating magnetization emerges along the x-axis with the sign of the signal determined by the sign of $\overline{J^{(1)}_0}$ (Fig. 3a,b), thereby distinguishing right and left enantiomers. The two curves in Fig. 3a sum to zero at all points, meaning that no signal will be observed in a racemic mixture.

 \begin{figure}[H]
 \includegraphics[width=0.9\textwidth]{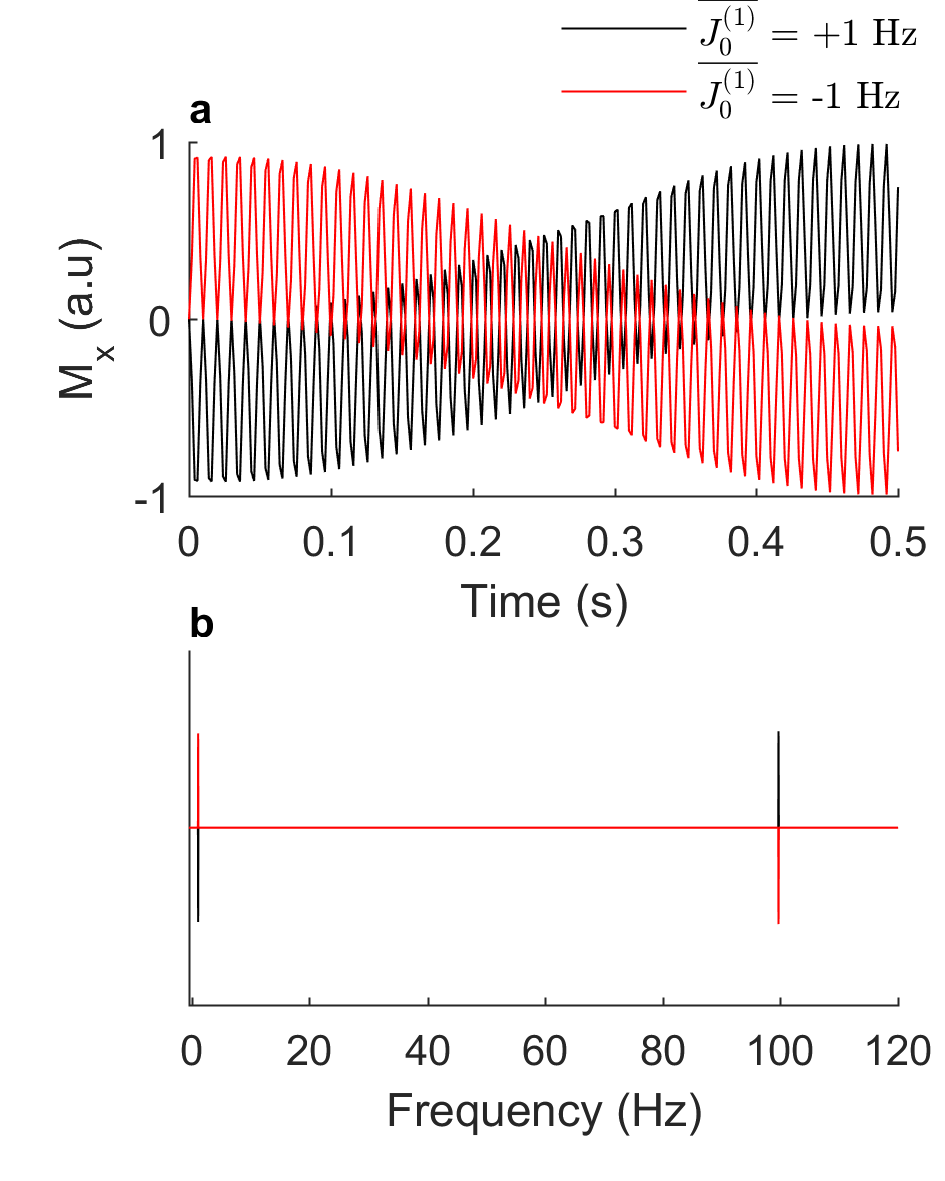}%
 \caption{\label{fig1}a)Predicted x-magnetzation for positive (black) and negative (red) $\overline{J^{(1)}_0}$ with a magnitude of 1 Hz, $J^{(0)}=100$ Hz, and a residual dipolar coupling of 0.7 Hz. High frequency oscillations result from coherences between the $|\pm1\rangle$ and $|\beta\rangle$ states while low frequency oscillations result from coherences between $|\pm1\rangle$ and $|\alpha\rangle$.  b) Fourier transform of (a) shows the phase relationship between peaks in the frequency spectrum for each enantiomer. 
 \label{results}
 }
 \end{figure}

The signal predicted by Eqn. \ref{signal} is proportional to $\overline{J^{(1)}_0}/J^{(0)}$, which scales the signal by the degree of orientation. It is useful to compare the amplitude of this chiral signal to the achiral signal that evolves along the y-axis. This achiral signal is the ``usual" zero-field NMR signal \cite{emagres} and is given by:

  \begin{equation}
\langle M_y(t)\rangle=-\frac{(\gamma^2_1-\gamma^2_2)(1+\left(\frac{\overline{J^{(1)}_0}}{J^{(0)}}\right)^2)}{N^2}\left[\cos(\omega_\alpha t)+\cos(\omega_\beta t)\right]
\label{}
\end{equation}

\noindent so that the relative amplitude of $M_x$ and $M_y$ is $\frac{4\gamma_1\gamma_2\frac{\overline{J^{(1)}_0}}{J^{(0)}}}{(\gamma_1^2-\gamma_2^2)(1+(\frac{\overline{J^{(1)}_0}}{J^{(0)}})^2)}$. For our experiment this ratio is $0.0108$, essentially in agreement with the exact value of $0.0106$ (Supplemental Material). Therefore we expect a signal reduced by a factor of $10^{-2}$ compared to standard zero-field NMR. A typical 100 $\mathrm{\mu}$L sample contains$\sim 10^{21}$ $^1$H-$^{13}$C pairs and, assuming a spherical geometry, the standard zero-field signal signal corresponds to a field of $\sim 50$ fT at a distance of 0.5 cm from the center of the sphere when prepolarized in 2 Tesla field at 300 K. Given a magnetometer sensitivity of $15$ fT/$\sqrt{\mathrm{Hz}}$ \cite{thomas}, and accounting for the $10^{-2}$ factor, we expect an acquisition time of approximately 5 minutes to achieve a signal to noise ratio greater than one.

We note that systematic errors may occur when the three axes (magnetizing field, electric field, and detector) are not orthogonal or when the pulse field is not collinear with the detector. In this case, some component of the large achiral signal will emerge along the detector axis. Such systematic errors could potentially overwhelm the desired chiral signal. However, we note that the desired signal is proportional to produce $J^{(1)}\cdot E$, whose sign can be changed by reversal of the electric field. The spurious achiral signal has no such dependence and thus may be cancelled by subtracting spectra acquired with reversed electric fields.

In conclusion, we demonstrate that molecular chirality is a directly observable property in NMR spectroscopy through antisymmetric terms in the spin-spin coupling Hamiltonian. Electric-field orientation in liquids provides a method to observe these terms, owing to the interplay between the pseudovector nature of the antisymmetric coupling and the polar vector nature of a molecular electric dipole. The apparent sign change of the spin coupling directly determines the sign of observed NMR signals, enabling chiral discrimination without adding additional chiral agents to the sample. Since the signal depends on pairwise spin-spin couplings, the spectrum gives local information and shows to what extent the local electronic structure is enantiomer-dependent. Furthermore, the observation of antisymmetric J-couplings will provide a new technique to search for molecular parity violation. This proposal combines previously established methods and in principle requires no new techniques. However, design constraints of our current zero-field NMR spectrometers do not allow the use of electric-field orientation cells. We therefore present this work as a guide to the design of future experiments.

\begin{acknowledgements}
The authors acknowledge professors Dmitry Budker, Mikhail Kozlov, Robert Harris, Alexander Pines, and Jamie Walls for helpful discussion. 
\end{acknowledgements}

\end{document}

%\preprint{}

%\title{Antisymmetric Couplings Enable Spectroscopic Resolution of Energy Equivalent Enantiomers in Nuclear Magnetic Resonance Spectroscopy}
\title{Antisymmetric Couplings Enable Direct Observation of Chirality in Nuclear Magnetic Resonance Spectroscopy}


\author{Jonathan P. King}
\email[]{jpking@berkeley.edu}
\affiliation{Department of Chemistry, University of California, Berkeley, California 94720, USA}
\affiliation{Materials Sciences Division, Lawrence Berkeley National Laboratory, Berkeley, California 94720, USA}
\author{Tobias F. Sjolander}
\affiliation{Department of Chemistry, University of California, Berkeley, California 94720, USA}
\affiliation{Materials Sciences Division, Lawrence Berkeley National Laboratory, Berkeley, California 94720, USA}
\author{John W. Blanchard}
%\author{Alexander P. Pines}
\affiliation{Department of Chemistry, University of California, Berkeley, California 94720, USA}
\affiliation{Materials Sciences Division, Lawrence Berkeley National Laboratory, Berkeley, California 94720, USA}
\affiliation{Helmholtz-Institut Mainz, Germany}

%\thanks{}


%\date{\today}

%\pacs{76.60.-k, 76.30.Mi, 82.56.Na}
%\keywords{}

%\maketitle

\section{Supplemental Material}
\subsection{Analytical Calculation}

The observed signal is given by:

\begin{equation}
\langle M_x(t)\rangle=\mathrm{Tr}\lbrace M^\dagger_x\rho(t)\rbrace=\sum_m\sum_nV_{nm}\rho_{mn}e^{-\frac{i}{\hbar}(E_n-E_m)t}
\label{}
\end{equation}

\noindent where $V_{nm}=\langle n|M_x|m\rangle$ and $\rho_{mn}=\langle m|\rho(0)|n\rangle$. Assuming no residual dipolar couplings and treating $\overline{J^{(1)}_0}$ as a first-order perturbation, the transition matrix elements are:
 
  \begin{equation}
V_{\pm1\alpha}=\frac{\sqrt{2}}{2N}\left[\gamma_1(\mp1-i\frac{\overline{J^{(1)}_0}}{J^{(0)}})+\gamma_2(\pm1- i\frac{\overline{J^{(1)}_0}}{J^{(0)}})\right]
\label{}
\end{equation}

and
  \begin{equation}
V_{\pm1\beta}=\frac{\sqrt{2}}{2N}\left[\gamma_1(1\pm i\frac{\overline{J^{(1)}_0}}{J^{(0)}})+\gamma_2(1\mp i\frac{\overline{J^{(1)}_0}}{J^{(0)}})\right].
\label{}
\end{equation}

Given the initial condition $\rho(0)=\frac{\gamma_1+\gamma_2}{2}(-I_{y,1}+I_{y,2})+\frac{\gamma_1-\gamma_2}{2}(-I_{y,1}-I_{y,2}),$ the coherence amplitudes are:

  \begin{equation}
\rho_{\pm1\alpha}=\frac{1}{N\sqrt{2}}\left[\gamma_1(-i\pm \frac{\overline{J^{(1)}_0}}{J^{(0)}})+\gamma_2(-i\mp \frac{\overline{J^{(1)}_0}}{J^{(0)}})\right]
\label{}
\end{equation}

and

  \begin{equation}
\rho_{\pm1\beta}=\frac{1}{N\sqrt{2}}\left[\gamma_1(\pm i-\frac{\overline{J^{(1)}_0}}{J^{(0)}})+\gamma_2(\mp i-\frac{\overline{J^{(1)}_0}}{J^{(0)}})\right],
\label{}
\end{equation}

and the observed signal is:

  \begin{equation}
\langle M_x(t)\rangle=\frac{4\gamma_1\gamma_2\overline{J^{(1)}_0}}{N^2J^{(0)}}\left[\cos(\omega_\alpha t)-\cos(\omega_\beta t)\right],
\label{}
\end{equation}

\noindent where $\omega_\alpha=\frac{E_\alpha-E_{\pm1}}{\hbar}$ and $\omega_\beta=\frac{E_\beta-E_{\pm1}}{\hbar}$.

We also consider the spurious, achrial signal that emerges along the y-axis. The relevant matrix elements of $M_y$ are:

  \begin{equation}
V^\prime_{\pm1\alpha}=-\frac{\sqrt{2}}{2N}\left[\gamma_1(\frac{\overline{J^{(1)}_0}}{J^{(0)}}\mp i)+\gamma_2(\frac{\overline{J^{(1)}_0}}{J^{(0)}}\pm i)\right]
\label{}
\end{equation}

and

  \begin{equation}.
V^\prime_{\pm1\beta}=\frac{\sqrt{2}}{2N}\left[\gamma_1(\frac{\overline{J^{(1)}_0}}{J^{(0)}}\mp i)-\gamma_2(\frac{\overline{J^{(1)}_0}}{J^{(0)}}\pm i)\right].
\label{}
\end{equation}

\noindent The coherence amplitudes are the same as before, so the signal is:

  \begin{equation}
\langle M_y(t)\rangle=-\frac{(\gamma^2_1-\gamma^2_2)(1+\left(\frac{\overline{J^{(1)}_0}}{J^{(0)}}\right)^2)}{N^2}\left[\cos(\omega_\alpha t)+\cos(\omega_\beta t)\right].
\label{}
\end{equation}

\subsection{Molecular Orientational Order Parameter}
Our proposed experiment involves liquid molecules that are partially oriented uniaxially by an applied electric field \cite{Buckingham1963,riley2000}, with rapid motion about the axis of orientation. Since the \textit{J}-coupling tensor is defined in the molecular reference frame, it is necessary to transform it to the reference frame of detection (laboratory frame). This is accomplished by the application of Wigner rotation operators:

\begin{equation}
J^{(q)}_{k'}=\sum_{k=-q}^qJ_{k}^{(q)}e^{-ik'\alpha}d_{k',k}^{(q)}(\beta)e^{-ik\gamma}
\end{equation}

\noindent where $(\alpha, \beta, \gamma)$ are Euler angles that define the orientation of molecular coordinate system in the detector reference frame. Rapid motion about the director axis averages over $\alpha$ so that only $k'=0$ terms are retained. The nonzero elements of the time averaged coupling tensor are then:

\begin{equation}
\overline{J^{(q)}_{0}}=\sum_{k'=-q}^qJ_{k'}^{(q)}\overline{d_{0,k}^{(q)}(\beta) e^{-ik\gamma}}.
\label{Jlab}
\end{equation}

\noindent The term in brackets may be regarded as a generalization of the order parameter for a uniaxially ordered system. Since $d^{(0)}_{0,0}=1$, the $q=0$ terms are unaffected by molecular motion and correspond to the well-known ``scalar coupling" component of the \textit{J}-coupling interaction commonly observed in liquid state NMR. The $q=1$ and $q=2$ terms depend on the degree of orientation and alignment, respectively. In our proposed experiment, orientation is generated by application of an electric field and alignment exists as an unavoidable byproduct of orientation. %However, owing to the second-order nature of the trigonometric terms $d^{(0)}$ as opposed to the first order terms in $d^{(1)}$ , the $q=2$ effects will be much smaller than the $q=1$ effects in the case of weak orientation. In any case, the presence of rank-2 terms does not interfere with the chiral discrimination, so we will ignore them in this example.

\subsection*{Simulations}
We take a system of two coupled spins, denoted $I_1$ and $I_2$, and define angular momentum operators as follows:

\begin{align}
&I_{1,x}  = \frac{1}{2}\left( \begin{array}{cccc}
0 & 1 & 0 & 0 \\
1 & 0 & 0 & 0 \\
0 & 0 & 0 & 1 \\
0 & 0 & 1 & 0 \end{array} \right),\quad I_{1,y} = \frac{1}{2}\I\left( \begin{array}{cccc}
0 & -1 & 0 & 0 \\
1 & 0 & 0 & 0 \\
0 & 0 & 0 & -1 \\
0 & 0 & 1 & 0 \end{array} \right), \nonumber \\
&I_{1,z} = \frac{1}{2}\left( \begin{array}{cccc}
1 & 0 & 0 & 0 \\
0 & -1 & 0 & 0 \\
0 & 0 & 1 & 0 \\
0 & 0 & 0 & -1 \end{array} \right)
\end{align}

\begin{align}
&I_{2,x}  = \frac{1}{2}\left( \begin{array}{cccc}
0 & 0 & 1 & 0 \\
0 & 0 & 0 & 1 \\
1 & 0 & 0 & 0 \\
0 & 1 & 0 & 0 \end{array} \right),\quad 
I_{2,y} = \frac{1}{2}\I\left( \begin{array}{cccc}
0 & 0 & -1 & 0 \\
0 & 0 & 0 & -1 \\
1 & 0 & 0 & 0 \\
0 & 1 & 0 & 0 \end{array} \right), \nonumber \\
&I_{2,z} = \frac{1}{2}\left( \begin{array}{cccc}
1 & 0 & 0 & 0 \\
0 & 1 & 0 & 0 \\
0 & 0 & -1 & 0 \\
0 & 0 & 0 & -1 \end{array} \right)
\end{align}

\noindent and define raising and lowering operators the usual way,
\begin{equation}I_{1,\pm}
I_{1,\pm}= I_{1,x} \pm \I I_{1,y}
\end{equation}

\begin{equation}
I_{2,\pm} = I_{2,x} \pm \I I_{2,y}.
\end{equation}

We assume that the system is first magnetized in a large magnetic field along the y-axis, which is switched suddenly to zero.  The density matrix of the system is then
\begin{equation}
\rho(0) = \frac{\E^{\frac{B_0\hbar}{kT}(\gamma_1 I_{1,y} + \gamma_2 I_{2,y})}}{\mathrm{Tr\lbrace\E^{\frac{B_0\hbar}{kT}(\gamma_1 I_{1,y} + \gamma_2 I_{2,y})}\rbrace}},
\end{equation}
assuming $kT$ $>>$ $B_0\gamma\hbar$, valid for any room temperature NMR system. We can then approximate the density matrix as a Taylor expansion truncated to first order
\begin{equation}
\rho(0) = \frac{1}{4}+\frac{B_0\hbar}{4kT}\left(\gamma_1 I_{1,y} + \gamma_2 I_{2,y}\right),
\end{equation}
The system is described by the following Hamiltonian
\begin{multline}
\frac{H}{\hbar} = 2\pi J^{(0)}\bm{I_1}\cdot\bm{I_2} + 2\pi\I \overline{J^{(1)}_0} (I_{1,+}I_{2,-}  - I_{1,-}I_{2,+}) + \\ 2\pi \overline{D^{(2)}_0}(-2I_{1,z}I_{2,z}+\frac{1}{2}(I_{1,+}I_{2,-}+I_{1,-}I_{2,+})),
\end{multline}
where $J^{(0)}$ is the scalar part of the spin-spin $J$-coupling, $\overline{J^{(1)}_0}$ is the time averaged antisymmetric part and $\overline{D^{(2)}_0}$ is the time averaged magnetic dipole-dipole interaction. Under $H$ the system evolves according to
\begin{equation}
\rho(t) = \E^{-\I \frac{H}{\hbar} t}\rho(0)\E^{\I \frac{H}{\hbar} t},
\end{equation}
which we simulate using discreet time steps
\begin{equation}
\rho(t+\Delta t) = \E^{-\I \frac{H}{\hbar} \Delta t}\rho(t)\E^{\I \frac{H}{\hbar} \Delta t}.
\end{equation}
The proposed experiment measures oscillating magnetization along the x-axis. We therefore want to calculate the expected time evolution of the x-magnetization, described by the operator $M_x = \gamma_1 I_{1,x} + \gamma_2 I_{2,x}$, where we again have ignored constants of proportionality.
\begin{equation}
\langle M_x(t)\rangle = \mathrm{Tr}\lbrace M_x^\dagger\rho(t)\rbrace = 0,
\end{equation}
however this expectation value is zero for all $t$, signifying that an initial state magnetized along y, does not yield an oscillating field along the x-axis.
We therefore modify our initial state by rotation about the x axis. Experimentally this would be implemented by a pulsed magnetic field to give
\begin{equation}
\rho^*(0) = \E^{-2\pi\I B_1 t(\gamma_1 I_{1,x} + \gamma_2 I_{2,x})}\rho(0)\E^{2\pi\I B_1 t(\gamma_1 I_{1,x} + \gamma_2 I_{2,x})},
\end{equation}
where $\rho^*$ denotes the modified state, $B_1$ is the amplitude of the pulse and $t$ its duration.
This initial condition does give a signal (the sign of which is determined by the sign of $\overline{J^{(1)}_0}_0$) along the x-axis.
\begin{equation}
\langle M_x(t)\rangle = \mathrm{Tr}\lbrace M_x^\dagger\rho^*(t)\rbrace,
\label{chiralSignal}
\end{equation}
This equation is evaluated numerically, with a time step $\Delta t$=0.002 s, and plotted in the main body of the paper (Figure 3a, with the fast Fourier transform plotted in Figure 3b.)
Maximum signal is achieved if $B_1t$ is chosen such that $\E^{2\pi\I B_1t\gamma_1} = -\E^{2\pi\I B_1t\gamma_2}$.
We took $\gamma_1 =$ 10.705 MHz/T  ($^{13}\mathrm{C}$) and $\gamma_2 =$ 42.576 MHz/T ($^{1}\mathrm{H}$), and let $B_1t\gamma_1 =$ $\pi$ and $B_1t\gamma_2 =$ 3.977$\pi$, making $\rho^*(0) \approx -\gamma_1 Iy + \gamma_2 I_{2,y}$. Note this expression is approximate because $\frac{\gamma_1}{\gamma_2}$ is not exactly equal to 4, and therefore a small amount of achiral signal will emerge along the z-axis. This signal will not be measured unless the pulse axis is misaligned from the detector axis. In this case it may be cancelled with the same electric field reversal discussed in the main text. 

We take representative values for the components of the \textit{J}-coupling tensor for a one-bond $^1$H-$^{13}$C system to be $J^{(0)} =   J_0^{(1)} =$ 100 Hz. We also assume that after orientational averaging $\overline{J^{(1)}_0} =$ 1 Hz, thus we assume the averaging scales the coupling by a factor 0.01. To estimate the electric field required to induce this degree of orientation we use the theory worked out in \cite{Buckingham1963}, which gives
\begin{equation}
\overline{J^{(1)}_0} = \frac{\mu E}{3kT}J^{(1)}_0,
\end{equation}
\noindent where $\mu$ is the electric dipole moment, $E$ is the electric field experienced by the sample, $k$ is the Boltzmann constant, and $T$ the temperature. The scaling is thus given by, $\mu E/3kT$. In polar liquids $E$ is related to the applied electric field as $E \approx [(\epsilon_r + 2)/3]E_{\mathrm{applied}}$, where $\epsilon_r$ is the static relative permitivity of the medium.
Polar organic molecules usually have dipole moments in the range 1-5 D \cite{riley2000} and static relative permitivities of 5-40. Assuming $\mu$ = 2.5 D, $\epsilon_r$ = 30 at a temperature of 298 K the electric field required to achieve $\mu E/3kT = 0.01$ is $E_{\mathrm{applied}} =$ 5 kV/cm. Fields between 5-40 kV/cm have been used previously in electric field aligned NMR experiments \cite{riley2000}.

Assuming a typical proton-carbon bond length of 1.092 \AA, these parameters will also yield a residual dipolar coupling of \cite{Buckingham1963}
\begin{equation}
\overline{D^{(2)}_0} = -\frac{1}{30}\left(\frac{\mu E}{kT} \right)^2 D^{(2)}_0 = 0.7 \mathrm{\,Hz}.
\end{equation}
This term has been included in the simulations, however the only effect is a small frequency shift of the signal, as seen in \cite{Blanchard2015}.

We note that the time evolution of states $\rho(0)$ and $\rho^*(0)$ also results in an oscillating magnetization along the y-axis. The corresponding signal is (ignoring second order energy shifts) independent of $\overline{J^{(1)}}$ and is in fact the usual signal detected in many zero-field NMR experiments \cite{emagres}. To estimate the magnitude of the signal predicted by eq.\ \ref{chiralSignal}, we compared the simulated peak amplitudes corresponding to the oscillating x- ($A_x$) and y- ($A_y$) magnetization respectively. The amplitudes were determined by fitting the Fourier transforms of $\mathrm{Tr}\lbrace M_x^\dagger\rho^*(t)\rbrace$, and $\mathrm{Tr}\lbrace M_y^\dagger\rho^*(t)\rbrace$ to complex Lorentzians, with the result
\begin{equation}
\frac{A_x}{Ay} = 0.0106,
\end{equation}
consistent with the analytical result of $0.0108$ in the main text.